\documentclass[fleqn,usenatbib,letters]{mnras}
\usepackage{newtxtext,newtxmath}
\usepackage[T1]{fontenc}
\usepackage{ae,aecompl}
\usepackage{graphicx}	% Including figure files

\usepackage{tikz}
\usepackage{color}

\newcommand{\Gaia}{\textit{Gaia}\ }

\title[Optical polarization of AGN with VLBI-\Gaia offsets]{Optical polarization properties of AGN with significant VLBI-\Gaia offsets}

\author[Y. Y. Kovalev et al.]{
Y.~Y.~Kovalev,$^{1,2,3}$\thanks{E-mail: yyk@asc.rssi.ru (YYK)}
D.~I.~Zobnina,$^{1}$
A.~V.~Plavin,$^{1,2}$
D.~Blinov$^{4,5,6}$
\\
$^1$Astro Space Center of Lebedev Physical Institute, Profsoyuznaya 86/32, 117997 Moscow, Russia\\
$^2$Moscow Institute of Physics and Technology, Institutsky per. 9, Dolgoprudny 141700, Russia\\
$^3$Max-Planck-Institut f\"ur Radioastronomie, Auf dem H\"ugel 69, 53121 Bonn, Germany\\
$^4$IESL \& Institute of Astrophysics, Foundation for Research and Technology-Hellas, 71110 Heraklion, Crete, Greece\\
$^5$University of Crete, Department of Physics, 71003 Herakleio, Crete, Greece\\
$^6$Astronomical Institute, St. Petersburg State University, Universitetsky pr.~28, Petrodvoretz, 198504 St. Petersburg, Russia
}

\date{Accepted 2020 January 8. Received 2019 December 23; in original form 2019 August 28}

\pubyear{2020}

\begin{document}
\label{firstpage}
\pagerange{\pageref{firstpage}--\pageref{lastpage}}
\maketitle

\begin{abstract}
Significant positional offsets of the value from 1~mas to more than 10~mas were found previously between radio (VLBI) and optical (\textit{Gaia}) positions of active galactic nuclei (AGN).
They happen preferentially parallel to the parsec-scale jet direction.
AGN with VLBI-to-\Gaia offsets pointed downstream the jet are found to have favourably higher optical polarization, as expected if extended optical jets dominate in the emission and shift the \Gaia centroid away from the physical nucleus of the source. Upstream offsets with the suggested domination of accretion disks manifest themselves through the observed low optical polarization.
Direction of linear optical polarization is confirmed to preferentially align with parsec-scale jets in AGN with dominant jets consistent with a toroidal magnetic field structure.
Our findings support the disk-jet interpretation of the observed positional offsets.
These results call on an intensification of AGN optical polarization monitoring programs in order to collect precious observational data. Taken together with the continued VLBI and \Gaia observations, they will allow researchers to reconstruct detailed models of the disk-jet system in AGN on parsec scales.
\end{abstract}

\begin{keywords}
galaxies: active~--
galaxies: jets~--
radio continuum: galaxies~--
astrometry~--
polarization
\end{keywords}

%%%%%%%%%%%%%%%%%%%%%%%%%%%%%%%%%%%%%%%%%%%%%%%%%%

%%%%%%%%%%%%%%%%% BODY OF PAPER %%%%%%%%%%%%%%%%%%

\section{Introduction}
\label{s:intro}

\begin{figure}
    \centering
    \includegraphics[width=0.9\columnwidth]{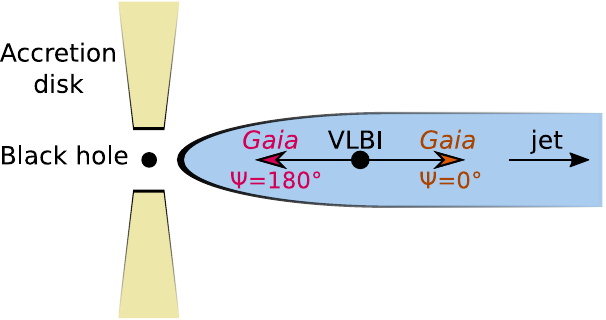}
    \caption{Diagram explaining the definition taken for the two preferred VLBI-\Gaia offset directions with respect to the parsec-scale jet: downstream $\Psi= 0\degr$ and upstream $\Psi=180\degr$.
    \label{f:diagram}
    }
\end{figure}

An analysis of the disk-jet system on parsec and sub-parsec scales is important for understanding nature of the central engine in active galactic nuclei. A new approach was suggested recently by utilizing VLBI-to-\Gaia offsets taken together with optical color information \citep{r:gaia1,r:gaia2,r:gaia3,r:gaia4,r:gaia5}. Using the Radio Fundamental Catalogue\footnote{\url{http://astrogeo.org/rfc/}} and \Gaia release~2 data \citep{r:GDR2cat}, offsets for 9\,\% of 9081 matched sources turned out to be significant \citep{r:gaia4}. The majority of them occur downstream or upstream the VLBI jet, see \autoref{f:diagram}. It has been shown that the downstream offsets are due to strong parsec-scale optical jets shifting the \Gaia centroid further away from the jet apex. Upstream VLBI-Gaia offsets occur when the accretion disk makes a major contribution to the optical emission while synchrotron opacity shifts radio away from the nucleus. 

In order to check this interpretation we introduce optical linear polarimetric data into the analysis. We expect that fractional polarization will allow us to distinguish the synchrotron radiation from the jet from the thermal emission from the accretion disk. 
We analyze properties of AGN with significant VLBI-\Gaia offsets forming an angle $\Psi$ with the jet direction (\autoref{f:diagram}).
The filtering threshold to select significant offsets is chosen to be $\sigma_\Psi<35\degr$ which roughly corresponds to a $2\sigma$ positional cutoff, see for details \citet{r:gaia5}. 
AGN are considered to have downstream VLBI-\Gaia offsets ($\Psi=0\degr$) if they have $\Psi\in(-45\degr,+45\degr)$, and upstream offsets ($\Psi=180\degr$) if observed $\Psi\in(180\degr-45\degr,180\degr+45\degr)$. The rest correspond to offsets with ``other'' direction.
Distribution of the angle $\Psi$ for different AGN classes is discussed by \citet{r:gaia5}.
%
%We do not discuss in this paper separately AGN with optical polarization data which have offsets not parallel with the jet direction since their total number is not high enough to deliver conclusive statistical results.
%
Most of the quasars in the analyzed sample have flat radio spectrum since they are selected on the VLBI-compact emission.

The letter is organized in the following way.
In \autoref{s:obsdata} we discuss optical data with linear polarization being used in the analysis, 
\autoref{s:pol-prop} compares electric vector positional angle (EVPA) of the optical polarization with parsec-scale jet direction as well as reports results on the fractional linear polarization observed from AGN of different types,
\autoref{s:discussion} and \ref{s:summary} present discussion and summary of the results.

\section{Optical polarization data in use}
\label{s:obsdata}

We collected archival optical polarimetric data from the literature. From \cite{Hutsemekers2005,Hutsemekers2018} we obtained single epoch measurements of 430 quasars. Their sample mostly consists of Broad Absorption Line, radio-loud and ``red'' quasars observed preferentially in the V band at FORS2 (VLT) and at EFOSC2 (3.6 m ESO telescope). Additionally, \cite{Hutsemekers2005} data set contains the R band and a broad band polarimetry collected from the literature. Effective wavelengths of the later measurements are close to that of the V and the R bands in all cases. The data quality cuts introduced by \cite{Hutsemekers2005,Hutsemekers2018} included $p \ge 0.6\%$, $\sigma_{\rm PA} \le 14\degr$ and $|b| \ge 30\degr$.

We also used data from two major optical polarization monitoring programs focused on blazars: Kanata \citep{Itoh2018} and RoboPol \citep{Blinov2019}. Using Kanata monitoring data for 27 and 37 sources in the R and V bands from \cite{Itoh2016}, we calculated the average fractional polarization of each source. To this end we employed the Maximum Likelihood method described by \citet{Blinov2016}. The advantage of this method is that it takes into account variability of polarization and measurements bias at the same time. Data of the RoboPol program were taken from \citet{Angelakis2016}, where the R band measurements for 158 sources are listed. The values of average fractional polarization for RoboPol data were calculated in the same way as we did it for Kanata. The average value of EVPA for each source in both programs was calculated as the position angle of the measurements centroid on the Q-U Stokes parameters plane. This method gives more persistent results with respect to calculation of median of the EVPA distribution in the case when it has the peak near the $180\degr - 0\degr$ transition.

In the case when a source was presented in both Kanata and RoboPol catalogues, the preference was given to latter one, because RoboPol monitoring was performed nearly simultaneously (2013-2014) with \Gaia operations.
In the case when both V and R bands polarization was given for the same source in the Kanata data, we used the R band values because of their typically higher signal-to-noise ratio. Since the wavelength dependence of polarization degree of AGN emission is rather weak in the optical band \citep[e.g.][]{Tommasi2001}, measurements in adjacent V and R bands are consistent within uncertainties in most of the cases. Therefore, we combined the collected data in a single set of average polarization parameters for 535 AGN. Of these objects 287 overlap with our sample of VLBI-Gaia offsets. Only 5 sources do not have significant polarization detected, and are excluded from our optical electric vector position angle analysis. 
Polarization fraction is considered to be equal to the reported detection limit for these 5 sources, as these upper limits are lower than all significant detections in the sample. 

\section{Optical polarization properties of AGN}
\label{s:pol-prop}

\begin{figure*}
\centering
\includegraphics[width=0.49\textwidth, trim=0.55cm 0.65cm 1cm 2cm,clip]{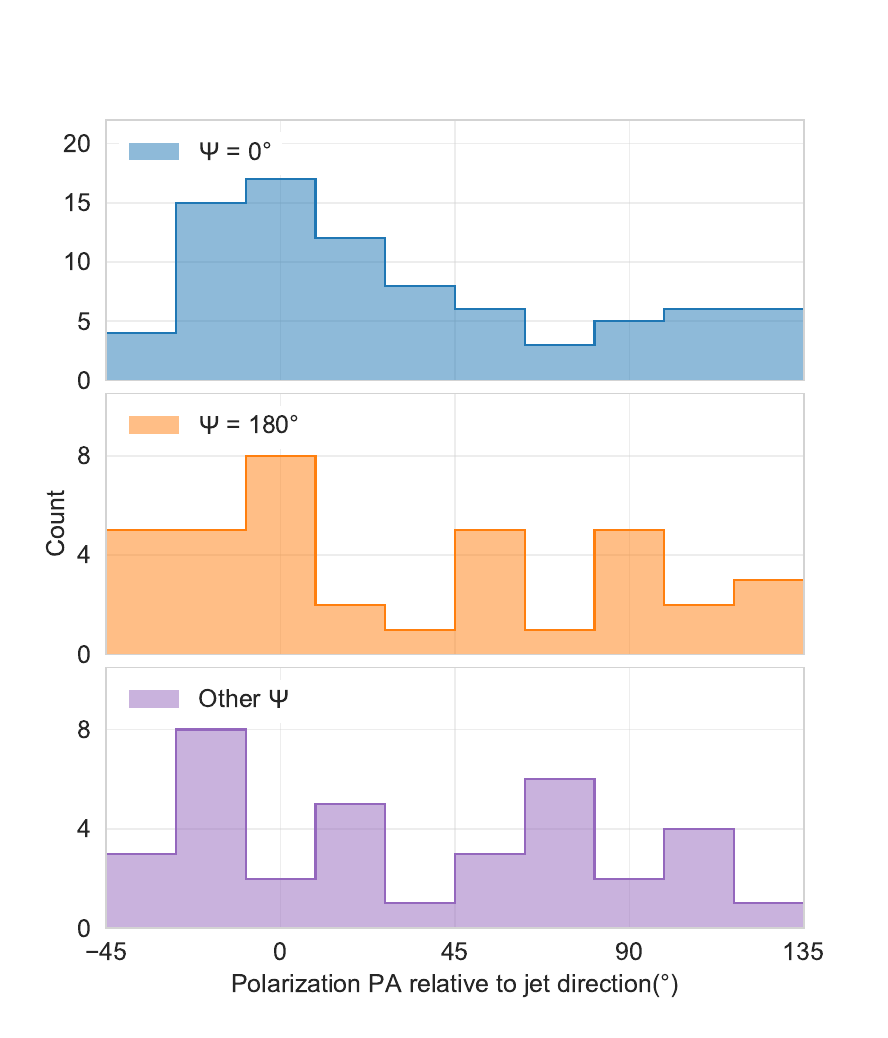}
\includegraphics[width=0.49\textwidth, trim=0.55cm 0.65cm 1cm 2cm,clip]{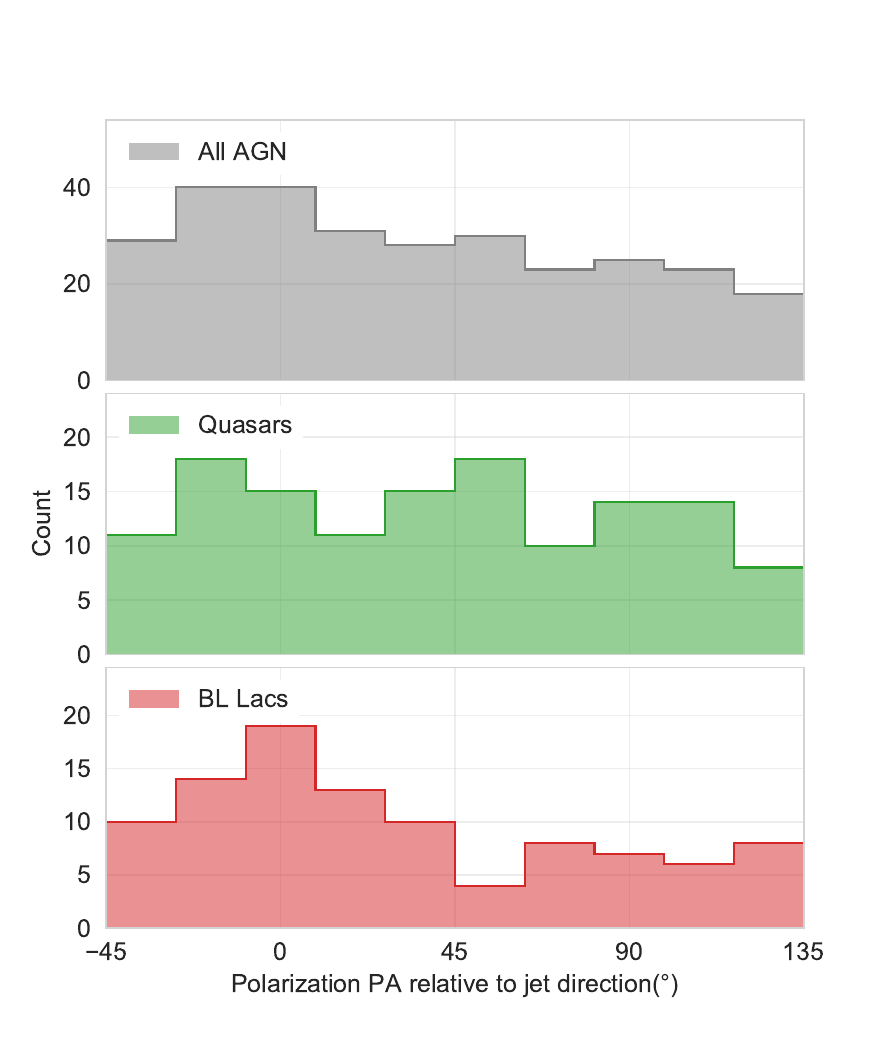}
\caption{
Distribution of the optical EVPA relative to the parsec-scale jet direction for AGN with different directions of significant VLBI-\Gaia offsets \textit{(left)}, 
for the full sample with measured optical polarization data as well as its sub-samples of quasars and BL~Lacs \textit{(right)}.
$\Psi=0\degr$ corresponds to downstream VLBI-\Gaia offsets, $\Psi=180\degr$ represents upstream ones.
See \autoref{t:sample_prop} for sample details.
\label{f:EVPA-jet}
}
\end{figure*}

\autoref{f:EVPA-jet} presents distribution of optical EVPA collected as discussed in \autoref{s:obsdata} relative to the parsec scale jet direction $\mathrm{P.A.}_\mathrm{jet}$ determined from VLBI images in \citet{r:gaia5}.
Significant peak at $\mathrm{EVPA}-\mathrm{P.A.}_\mathrm{jet}=0\degr$ is observed in the full sample as well as for BL~Lacs: see \autoref{f:EVPA-jet} for illustration and \autoref{t:sample_prop} for significance estimates.
Moreover, among AGN with significant VLBI-\Gaia offsets those with downstream offsets, i.e. $\Psi=0\degr$, give the most pronounced peak of $\mathrm{EVPA}-\mathrm{P.A.}_\mathrm{jet}$ distribution at $0\degr$. This sample is expected to contain AGN with optical emission predominantly coming from the jet \citep{r:gaia5}. 
Significance of the $0\degr$-peak in the distributions is assessed using the bootstrap method with 10000 random realizations \citep[see for details][]{Press:2007:NRE:1403886}. We count sources with $-30\degr<\mathrm{EVPA}<30\degr$ for each realization. The fraction of realizations providing counts in this range higher than expected for uniform distribution is taken as the probability of the peak significance.
We note that the preference of EVPA to be aligned with the jet has been reported previously in a number of publications but did not result in a conclusive outcome \citep[e.g.][]{RS85,Lister2000,r:algaba11,Hovatta2016,Angelakis2017}.

\begin{table}
\caption{Sample characteristics.
Columns are as follows:
(1) Sample;
(2) number of AGN;
(3) significance probability of the peak at $0\degr$ in the distribution of the optical polarization $\mathrm{EVPA}-\mathrm{PA}_\mathrm{jet}$ value (\autoref{f:EVPA-jet});
(4) median optical linear polarization fraction and its 68\% uncertainty (\autoref{f:optpolfrac}, \ref{f:pd_median_summary}).
\label{t:sample_prop}
}
\centering
\begin{tabular}{lrcc}
\hline\hline\noalign{\smallskip}
Sample    & $N$ & $0\degr$-peak significance probability &   $p_\mathrm{med}$~(\%)  \\
(1)             & (2) & (3)    &  (4)          \\
\hline
All AGN         & 287 & ~0.999 &  $3.7_{-0.2}^{+0.2}$  \\
\hline
Quasars         & 134 & ~0.789 &  $2.5_{-0.5}^{+0.6}$  \\
BL~Lacs         & 99  & >0.999 &  $8.1_{-0.6}^{+0.3}$  \\
\hline
$\Psi=0\degr$   & 82  & >0.999 &  $4.7_{-0.8}^{+0.4}$  \\
$\Psi=180\degr$ & 37  & ~0.932 &  $1.2_{-0.3}^{+0.2}$  \\
Other $\Psi$    & 35  & ~0.887 &  $3.7_{-1.3}^{+1.0}$  \\
\hline
\end{tabular}
\begin{flushleft}
Note. The samples of all AGN, quasars and BL~Lacs represent AGN which have optical polarization information (\autoref{s:obsdata}) and VLBI-\Gaia counterparts \citep{r:gaia4,r:gaia5} at any level of the VLBI-\Gaia offset significance. The ``All AGN'' sample includes quasars and BL~Lacs as well as Seyfert and radio galaxies. The samples selected on the $\Psi$ value are drawn from the ``All AGN'' sample after applying the filter on the $\sigma_\Psi$ value, as discussed in \autoref{s:intro}. The method to estimate the probability of the peak significance (3) is discussed in \autoref{s:pol-prop}.
\end{flushleft}
\end{table}

\begin{figure*}
\centering
\includegraphics[width=0.49\textwidth]{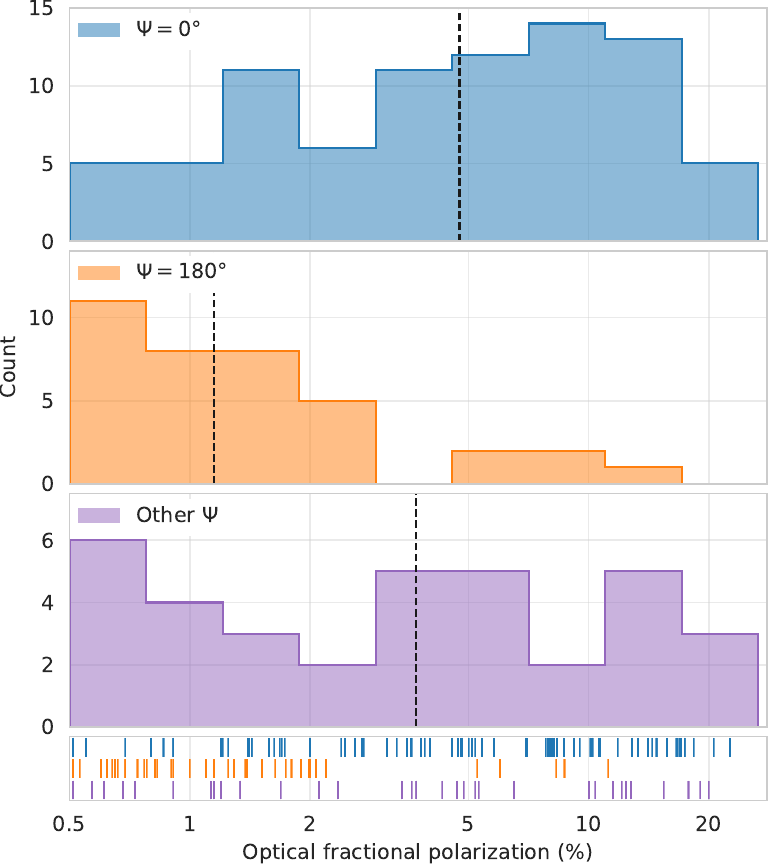}
\includegraphics[width=0.49\textwidth]{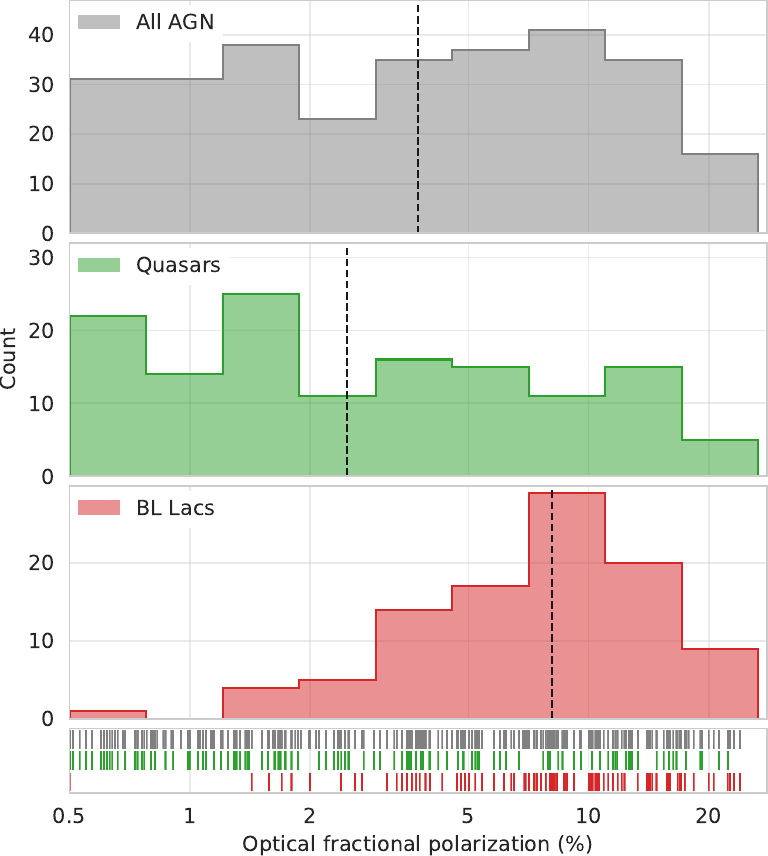}
\caption{
Distribution of the fractional optical linear polarization for 
objects with different directions of significant VLBI-\Gaia offsets \textit{(left)},
for the full sample with measured optical polarization data as well as its sub-samples of quasars and BL~Lacs \textit{(right)}.
See \autoref{t:sample_prop} for sample details.
The sticks on the bottom present fractional polarization values for individual sample members. The vertical dashed lines represent medians of corresponding subsamples.
Note that the horizontal axis is logarithmic, and values less than $0.5\%$ are shown as $0.5\%$.
\label{f:optpolfrac}
}
\end{figure*}

\begin{figure}
\centering
\includegraphics[width=\columnwidth,trim=0cm 0.5cm 0cm 0cm]{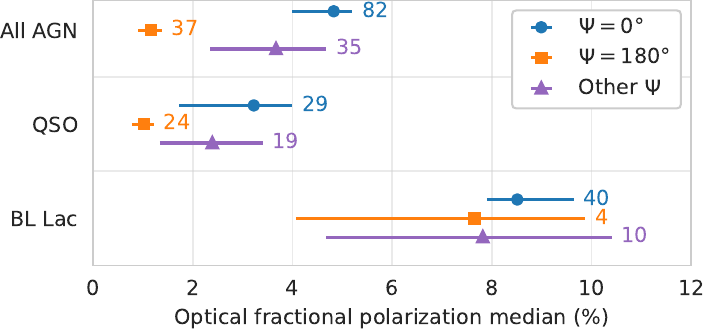}
\caption{
Median polarization fraction for quasars, BL Lac objects and all AGN
which have significant VLBI-\Gaia offsets ($\sigma_\Psi<35\degr$, see for selection details \autoref{s:intro}). Errorbars show $68\%$ uncertainties.
Objects are grouped by the offset direction $\Psi$. The numbers shown in the plot represent sizes of corresponding sub-samples.
\label{f:pd_median_summary}
}
\end{figure}

The $\Psi=180\degr$ AGN are found to have the typical level of fractional linear optical polarization several times lower than the $\Psi=0\degr$ case (\autoref{f:optpolfrac}, \autoref{f:pd_median_summary}, \autoref{t:sample_prop}). Namely, the fractional polarization median for all AGN with upstream VLBI-\Gaia offsets is found to be lower than that for downstream offsets with the significance probability $>0.999$.
If BL~Lacs are AGN with the highest dominance of synchrotron emitting jets in their radiation \citep{r:UP95}, most of them should show downstream VLBI-\Gaia offsets \citep[see for details][]{r:gaia5} and the highest polarization fraction. The later was known before \citep{Impey1990} and is confirmed by us here.
At the same time, BL~Lacs do not solely drive the found difference between optical polarization fraction for upstream and downstrem offsets in the ``All AGN'' sample. This  difference is confirmed with the probability of significance $0.998$ if we analyze the sample of quasars only.
We assess uncertainties of the medians as well as the significance probability of the difference between them by the same bootstrap approach as described above. 
The probability of medians for two subsamples being $p_\mathrm{1,med}>p_\mathrm{2,med}$ is calculated as
the fraction of realizations for which this inequality holds.
It would be interesting to perform a comparison of optical polarization properties of objects with shorter and longer VLBI-\Gaia offsets. Unfortunately, we do not find any significant dependence due to small sample sizes.

It is worth noting that the collected list of sources with archival polarimetric data (\autoref{s:obsdata}) may be biased towards objects with higher degree of optical polarization. For this reason we also repeated our analysis for the $\gamma$-ray loud  and the $\gamma$-ray quiet samples of RoboPol separately. These samples were created using strict, well-defined criteria described by \cite{Pavlidou14} and present unbiased populations of blazars that belong to corresponding classes. This analysis confirmed our findings, even though the found differences are less significant due to smaller sample sizes.
%
%%%Dropped to make the statistical analysis discussion more simple:
%Interestingly, additional expected AGN properties were confirmed.
%Gamma-ray loud AGN typically have higher Doppler-boosting \citep[e.g.][]{MOJAVE_Fermi1,MOJAVE_Fermi2} and due to that higher jet dominance and higher fractional optical polarization \citep[e.g.][]{Angelakis2016}. Not surprizingly, 63\,\% of them belong to the $\Psi=0\degr$ type and only 16\,\% to the $\Psi=180\degr$ one. An opposite situation is observed for $\gamma$-ray quiet AGN from the RoboPol sample: 11\,\% show $\Psi=0\degr$ and 55\,\% exhibit $\Psi=180\degr$ offsets.

\section{Discussion}
\label{s:discussion}

The optical emission from AGN is composed of several components, of which the main contributors are the jet, the accretion disc, and the broad line region. The emission mechanisms and the physical conditions of these components are very different, which in turn is reflected in the polarization characteristics of the emitted light. Moreover, if  several components have comparable levels of linearly polarized flux density but different EVPA, it results in depolarization of the measured integrated polarized signal and affects the EVPA. This could be observed in our data (\autoref{f:optpolfrac}), if levels of polarized emission produced by the disk and the jet are comparable. The total emission of the jet itself might get depolarized as well.
%which might happen for the $\Psi=180\degr$ case.

The synchrotron radiation from relativistic jets is known to be highly polarized from both the theoretical \citep[e.g.][]{Pacholczyk1970,Ginzburg1979} and observational \citep[e.g.][]{Angelakis2016,MOJAVE_XV} perspective. The theoretical limit is about 80\,\% for optically thin synchrotron emission in a uniform magnetic field. The fractional polarized emission from jets varies from several per~cent, typical for opaque radiation of the core, up to several tens of per~cent for the optically thin jet regions as observed in radio \citep{r:pus17}. The opaque radio core typically dominates the total and linearly polarized radiation at parsec scales in AGN \citep[e.g.][]{2cmPaperIV,Hodge18}. The maximum observed polarization in the optical band is 45\,\% \citep{Mead1990}. 
%The Faraday rotation does not affect the EVPA significantly in the optical band due to the very short wavelength. This means that knowing the optical EVPA and assuming the optically thin regime, we can get an idea of the magnetic field orientation in the jet region which dominates in polarization.

The thermal radiation from the accretion disk becomes polarized on passing through its atmosphere  due to scattering. Theoretical models predict a general trend of rising fractional polarization with increasing viewing angle of the disk rotation axis, $\theta$, up to 10-20\,\% and even higher.
For example, rather simplistic models considering only Rayleigh scattering predict a maximum polarization of 11.7\,\% for $\theta=90\degr$ \citep{Chandrasekhar, Sobolev}. 
\citet{1996MNRAS.282..965A} also took magnetic fields into consideration.
\citet{Beloborodov} took into account scattering on a mildly relativistic disk wind.
Further complications are introduced by the electrons and dust located in the equatorial plane beyond the disk, in the torus and in the polar regions \citep{Goosmann,Wolf}.
However, the sample analysed in this paper is highly dominated by AGN with small viewing angles $\theta\approx5\degr$, where the fractional polarization of the disk emission does not exceed 1\,\% for all models mentioned above, while the EVPA orientation with respect to the disc axis depends on the model.

The study of the relationship between the VLBI jet direction and the optical EVPA (\autoref{s:pol-prop}) confirms that the position angle of linear polarization is perferentially aligned with the parsec-scale jet direction.
This is found most clearly for AGN with downstream offsets, which are expected to have dominant radiation from their jets in the optical domain \citep{r:gaia5}.
It is safe to assume that the dominant component of the linearly polarized optical emission comes from the jet, which emits optically thin synchrotron radiation. A toroidal magnetic field can produce such polarization angles \citep[e.g.][]{Lyutikov2005}.

The comparison of the VLBI-\Gaia positional shift and the optical polarisation fraction (\autoref{s:pol-prop}) strongly supports our hypothesis: most of the  observed VLBI-\Gaia offsets happen due to the physical properties of the disk-jet system. 
The median value of fractional polarization for the AGN with upstream offsets is found to be the lowest among \textit{all} samples discussed, with and without filtering on the offset direction (\autoref{t:sample_prop}, \autoref{f:pd_median_summary}).
This confirms that accretion disks dominate or at least constitute a significant fraction of the optical emission of AGN with $\Psi=180\degr$ while relativistic jets define the observed polarization for $\Psi=0\degr$.
These results show that optical polarization data provide a critical additional source of information, complementing VLBI and \Gaia measurements, which will allow us to separate and study contributions from the accretion disk and relativistic jet in AGN emission \citep[see also][]{Li2019}.

Note that dust can affect the centroid position. 
As was shown by \cite{r:gaia5} for Seyfert galaxies, the central region of AGN can be obscured by a torus, producing downstream offsets for nearby sources observed at large angles to their jets. We have only 45 Seyfert and radio galaxies in the sample of 287 AGN analyzed, which is  dominated by blazars.
The dust extinction can also affect the unbeamed emission from the host galaxy. However, the majority of blazars are hosted by giant elliptical galaxies which show a low level of extinction \citep{benn98}. Additionally, the host galaxy extinction cannot produce the clear alignment found between the offsets and parsec-scale jet direction \citep{r:gaia2,r:gaia5} as well as the non-linear offset jitter \citep{r:gaia4}.
The host galaxy influence should vanish with redshift; however, we see offsets for quasars in both directions up to high redshift values \citep{r:gaia5} as well as the fractional polarization difference predicted for disk-jet systems (\autoref{f:pd_median_summary}).

\section{Summary}
\label{s:summary}

We have found that AGN with VLBI-\Gaia offsets upstream in the jet have significantly lower fractional linear optical polarization than the downstream ones. This is indeed expected within the proposed scenario: optical emission with the centroid close to the AGN nucleus corresponds to the dominance of an accretion disk and if the \Gaia centroid shifts down the parsec-scale jet, the jet emission dominates.
This outcome can be easily understood by the difference in their radiation mechanisms.
Scattered thermal emission from accretion disks is significantly less polarized than the synchrotron emission from the jets. 
Many AGN with jet dominance in their optical radiation show an optical polarization direction aligned with the jet direction, which confirms and extends the results reported by \cite{RS85} and supports models with a toroidal magnetic field \citep[e.g.][]{Lyutikov2005}.

The results of the current work, together with the studies by \citet{r:gaia5} and \citet{r:gaia4}, unambiguously demonstrate that there is a large population of AGN with an optical structure at milli-arcsecond scales that can be probed using \Gaia in combination with VLBI.
This opens an entirely new window for AGN studies in the era of multi-messenger astronomy, and urges us and other groups to start a systematic survey of the optical polarization of AGN with significant VLBI-\Gaia offsets. If performed during the \Gaia mission life time it will secure the most accurate contemporaneous measurements of VLBI-\Gaia offsets. Thereby, we will be able to discriminate between the disc and jet dominated AGN, study their polarization properties on a statistical basis, and better understand the structure and physics of disk-jet systems. An ultimate test of the conclusions presented here could be made by separating flares generated in disks and jets on the basis of their observed fractional polarization and comparing with the VLBI-\Gaia offset variations. Moreover, given that \Gaia operates in a scanning mode, observes each target every 25 days on average \citep{r:Gaia}, and that the monitoring positional data will be available in future data releases, we encourage polarimetric and multiband photometric monitoring programs at R and B bands. Table~2 (electronic only) presents data on 1059 AGN with VLBI-\Gaia offsets filtered on $\sigma_\Psi<35\degr$ and \Gaia V magnitude brighter than 19.0 for selecting promising targets for monitoring. Attention should be paid to \Gaia alerts on AGN flares\footnote{\url{http://gsaweb.ast.cam.ac.uk/alerts/alertsindex}} since there is a prediction and a preliminary confirmation of offset variations during the flares \citep{r:gaia4,r:gaia5}. We foresee that such observations can be used to resolve a number of open questions in AGN physics including the following: (i) where do orphan optical flares originate and what is the mechanism producing them; (ii) are there systematic differences between flares in disks and jets and how do they evolve, (iii) what is the mechanism behind optical polarization position angle rotations and where do they happen?

\section*{Acknowledgements}

We thank V.~Afanasiev, S.~Anton, I.~Browne, L.~Petrov, R.~Porcas, E.~Ros, S.~Wagner, and the referee E.~Perlman for discussions and suggestions which helped to improve the manuscript.
This project is supported by the Russian Science Foundation grant 16-12-10481.
This research has made use of NASA's Astrophysics Data System. DB acknowledges support from the European Research Council (ERC) under the European Union's Horizon 2020 research and innovation program under grant agreement No.~77128.

%\bibliographystyle{mnras}
%\bibliography{vlbi_gaia_optpol}

% Don't change these lines
\bsp	% typesetting comment
\label{lastpage}
\end{document}